\def\year{2019}\relax
\documentclass[letterpaper]{article} 
\usepackage{aaai19}  
\usepackage{times}  
\usepackage{helvet}  
\usepackage[table]{xcolor}
\usepackage{amsmath}
\usepackage{amsthm}
\usepackage{times}
\usepackage{graphicx}
\usepackage{latexsym}
\usepackage{multirow}
\usepackage{tikz}
\usepackage{colortbl}
\usepackage{amscd,amsmath,graphicx,latexsym}
\usepackage{pgfplots}
\usepackage{float}
\usepackage{tablefootnote}
\usepackage{hyperref}
\usepackage{subfigure}
\usepackage{amssymb}
\usepackage{color}
\usepackage{amsmath}
\usepackage{graphicx}
\usepackage{url}
\newtheorem{example}{Example}
\usepackage{courier}  
\usepackage{url}  
\usepackage{graphicx}  
\begin{document}
%
\title{Using deceased-donor kidneys to initiate chains of living donor kidney paired donations: algorithms and experimentation.}
\author{
Cristina Cornelio \\
IBM Research \\
ccornel@us.ibm.com\\
\And
Lucrezia Furian \\
University of Padova	 \\
lucrezia.furian@unipd.it\\
\And
Antonio Nicolo' \\
University of Padova \\
antonio.nicolo@unipd.it\\
\And
Francesca Rossi \\
IBM Research\\ 
University of Padova \\
frossi@math.unipd.it\\
}

\maketitle
\begin{abstract}
We design a flexible algorithm that exploits deceased donor kidneys to initiate chains of living donor kidney paired donations, combining deceased and living donor allocation mechanisms to improve the quantity and quality of kidney transplants. The advantages of this approach have been measured using retrospective data on the pool of donor/recipient incompatible and desensitized pairs at the Padua University Hospital, the largest center for living donor kidney transplants in Italy. The experiments show a remarkable improvement on the number of patients with incompatible donor who could be transplanted, a decrease in the number of desensitization procedures, and an increase in the number of UT patients (that is, patients unlikely to be transplanted for immunological reasons) in the waiting list who could receive an organ.

\end{abstract}

\section{Introduction}

Living donor kidney transplantation is the most promising solution for closing the gap between organ demand and supply. Despite growing efforts to implement this option for patients with end-stage renal disease (ESRD), the percentage of living donations has been decreasing in the United States from 50.1\% in 2000 to 38.2\% in 2017 of the total number of kidney transplants \cite{OPTN}. Figures are different in the European countries, where the expansion of living transplantation programs is still underway: in Italy there were 310 kidney transplants from living donor in 2017, less than one fifth of the total number of kidney transplants. 

Patients may have a willing living donor, but they cannot receive her organ due to blood or tissue type incompatibility.  Some ABO-types  (blood) incompatibilities between donor and recipient can be resolved through desensitization techniques (which are costly and may have side-effect of the health of the patients), but in many other instances, and especially in case of patients with circulating human leukocyte antigen (HLA) antibodies directed against their willing donors, incompatibility cannot be overcome. 

A welfare enhancing option for incompatible donor/recipient pairs is to participate to kidney paired exchange (KPE) programs that favor donors' exchanges among incompatible pairs. A relevant constraint in designing the algorithms to maximize the number of  transplants in KPE programs, is the simultaneity of the exchanges among pairs. It follows that there are feasibility constraints over the length of these cycles, because three-way cycles already represent a logistic challenge for most of the transplant centres.

Unfortunately, these programs are not fully and uniformly developed among Europe, and countries such as Spain, France, Italy, Czech Republic, Austria, Belgium, Switzerland, Poland and Scandinavia have just started these programs or are striving to implement one. 
In these countries, few kidney paired exchanges have been performed, mainly due to the low number of patients enrolled in such programs. In Italy, for instance, there are currently only 39 donor/recipient pairs enrolled in the National Kidney Paired Donation program. It is quite unlikely to find two-way or three-way cycles of exchanges among such a few pairs.  

A suitable option to increase the number of transplants is the inclusion of altruistic donors to initiate chains of {\it kidney paired donations (KPD)}. Altruistic donors enabled further expansion of this practice, because the availability of a non-directed living donor kidney without a designated recipient increases the number of potential matches. Moreover, chains starting from a deceased donor (DD) are not constrained by simultaneous exchanges, because every patient involved in a KPD chain will receive an organ before her donor donates to the following patient in the chain. Therefore, in case of a break of a chain, it never happens that a patient does not receive an organ but her donor has already donated a kidney to another patient. However, the uncertainty among both the transplantation community and the public opinion with regard to the intention, motivation, and legitimacy of such donors represents a constraint to a large development of this option in many countries.

In the present study, we consider the use of deceased donors to initiate chains of KPD.  We propose to design a new program that combines the benefit of a KPD program and of a deceased donors' program, and propose algorithms to implement it.  Moreover, we  analyze and quantify its potential benefits by using retrospective data of the Kidney and Pancreas Transplant Unit at Padua University Hospital, one of the largest kidney transplant centre in Italy.

\subsection{Related works}

This paper contributes to the kidney exchange literature initiated by \cite{Delmon2004,delmonicoAJT,kidney_exchange_roth,Su_Zenios_Nephrology} and developed successively by \cite{pairwise_kidney_exchange_roth,kidney_europe,Roth05_clearinghouse,Unver_Dynamic,online_Sandholm,free_riding}. 
These algorithms have been further analyzed: studying the problem of transplantation failures due to rejected matching \cite{reni_cycles_chains_failure}; 
focusing on particular classes of patients (e.g. patients with a very low probability of transplantation) \cite{reni_fairness}; 
analyzing the computational complexity of the problem as ILP \cite{reni_cicli};
providing an analysis of the efficacy of altruistic-donor chains \cite{reni_chain}; 
studying the problem of having long chains of kidney exchanges \cite{long_chains_efficient,long_chains};
providing incentives to compatible pairs to participate in these programs \cite{NicoloRA,NicoloAgeBased,reni_compatible_pairs};
using altruistic donors to initiate chains of donation \cite{sonunvaltruistic,simultaneous_altruistic,Manlove2014,altruistic_chains}.

The ethical implications of the utilization of deceased-donors grafts to start chains of donations has been analyzed previously \cite{doctors5}. In \cite{doctors6} an approach for using deceased-donors to start donation chains is proposed, giving priority in the waiting list to patients that have a donor (a chain starts only if one of these particular patients is selected from the deceased-donor waiting list).
A similar approach that modify the priority of patients in the waiting list (with the emission of vouchers) is developed in \cite{doctors4}.
Another similar work \cite{RSUAJTrans} adopts list exchange procedures according to which a living incompatible donor provides a kidney to a candidate on the deceased-donor wait-list and in return his intended recipient receives a `priority on the deceased-donor wait-list.
We differentiate from these works mainly in two ways: we do not modify the waiting list (that is handled with a separate standard algorithm); we give priority to the UT patients (patients with unlikely transplantability).
Also in \cite{doctors10} the authors consider deceased-donors to start chains of donation: this approach can be better described as a ``List Exchange procedure''. Contrary to our work, a pair donates before his recipient receives an organ from, and there is a degree of uncertainty of the prioritization in the waiting list (mostly in case of highly sensitized patients).

\section{The Algorithm}



Given a compatibility graph that includes all patients, donors, and organs (to be precisely described later), our algorithm looks for cycles or chains in this graph, and it is triggered by either the arrival of a new patient/donor pair (in this case we look for a cycle, via the {\it cycle detection} procedure), or of a new deceased-donor kidney (in this case we look for a chain, via the {\it chain detection} procedure, and we release the donor of the last chain element to the deceased donor waiting list). Both procedures solve an optimization problem on a graph structure, and aim at maximizing the number of transplants.

\begin{figure}[h!]
\centering
\includegraphics[scale=0.4]{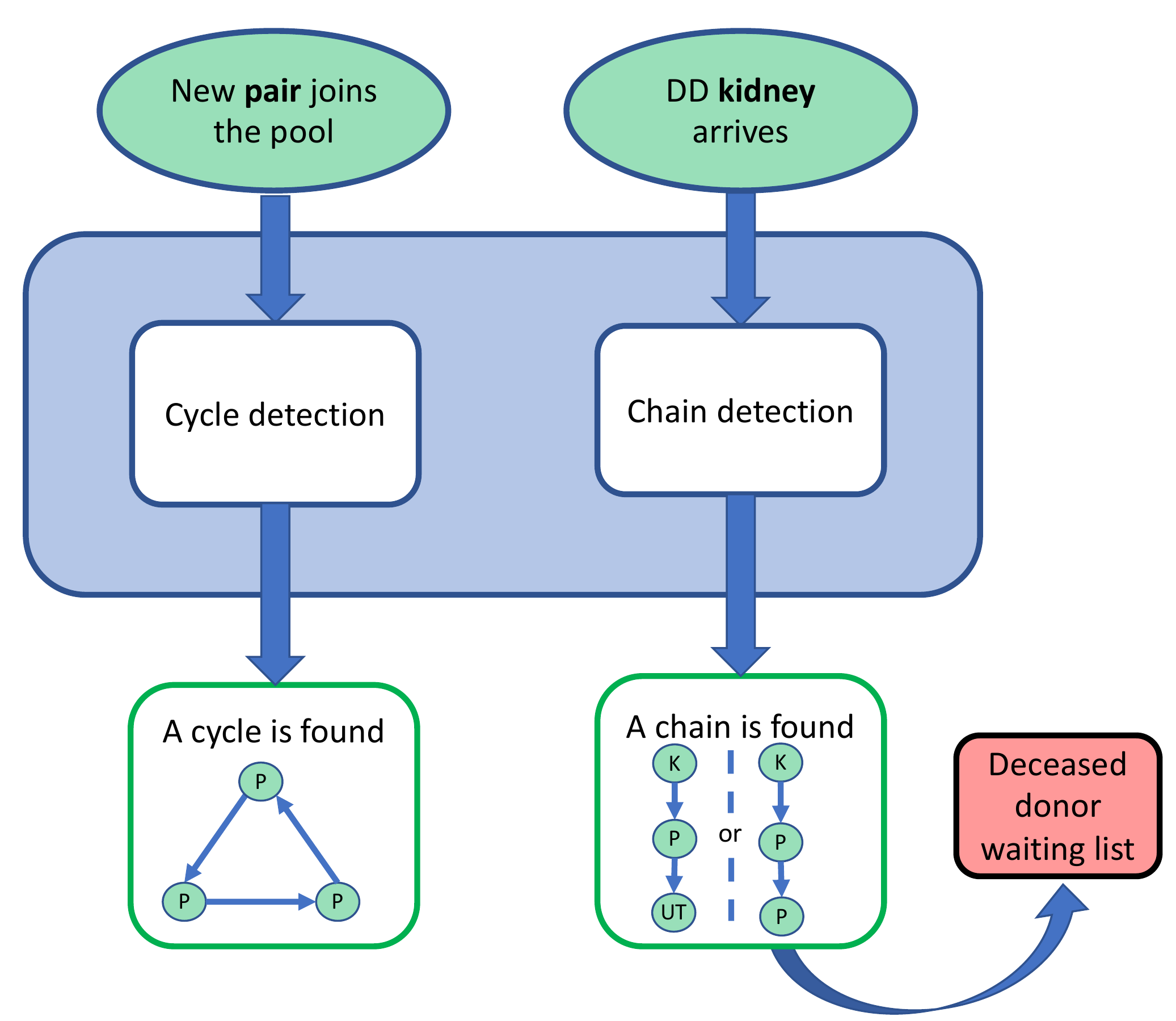}
\caption{The algorithm.}
\label{fig:algorithm}
\end{figure}

An overview of the algorithm is depicted in Figure \ref{fig:algorithm}.

\subsection{Patient pool}
We now give the details of our scenario, where we consider the following types of patients:
\begin{itemize}
\item {\it UT patients}: these are patients with {\it unlikely transplantability}, that is, patients that are difficult to match with donors since they are highly immunized. Usually this type of patients are at their second or third transplant.
\item {\it NT pairs donor/patient}: these are pairs that are not compatible between each other and therefore cannot be transplanted between themselves.
\item {\it DS pairs donor/patient}: these are pairs that are not compatible between each other but they can be transplanted between themselves with desensitization. Desensitization is a procedure that solve light incompatibility (like the blood type or the RH type) but it induces stress in the body of the patient, who is already week from the dialysis. It also has a high cost from a financial point of view, so both for the patient and the hospital would be better to be avoided.
\end{itemize}

We define our pool of patients as a set that can include UT patient, NT pairs, DS pairs, or optimal kidneys from deceased donors\footnote{In the {\it Experiments} section we will describe further restrictions on the pool composition, differently for each experimental setting.}. The composition of the pool varies with the time: given each time stamp $t$ we denote the pool at that particular moment as $P(t)$. $P(t)$ contains all the UT patients, NT pairs, DS pairs that are in the deceased donor waiting list at that particular time $t$, and the kidneys arrived in the pool at time $t$. We are not interested in all the time stamps $t$ but only those at which the pool composition change. These particular moments happen when a new pair join the deceased-donor waiting list or when a new deceased-donor kidney arrives.

\subsection{The compatibility graph}

We represent each element of the pool (each pair NT, pair DS, patient UT, or kidney) as a node of a {\it compatibility graph}, which is a directed, possibly cyclic, graph. In the graph there are three different types of nodes: nodes representing to donor/patient pairs (either UT pairs or DS pairs); kidney nodes that correspond to a deceased donor kidney; and UT patient nodes that correspond to recipients of unlikely transplantability. Edges in the graph instead represent the compatibility between two nodes: we add an edge $(X, Y)$ from node $X$ to node $Y$ only if the donor/kidney of node $X$ is compatible with the patient of node $Y$. Nodes representing kidneys only have outgoing edges and UT patients nodes have only incoming edges. Pair nodes may have both outgoing and incoming edges. Therefore we partition the set of nodes $N$ in three sets: $P$ the set of the pair nodes, $UT$ the set of immunized patient and $K$ the set of kidney. Given a node $X$, we call $E_X$ the set of its edges, partitioned in two subsets: the incoming edges $E_X^-$ and the outgoing edges  $E_X^+$.


An edge can be {\it active} or {\it inactive}: to each edge $e$ is associated a value $v_e$ in $\{0,1\}$ such that $v_e = 1$ if the edge is active and $0$ if inactive. The semantics of the active/inactive edges is that the edges represents possible transplants, and the active ones represent the transplants that are performed in practice at that time stamp $t$.

\subsection{Constraints}

A set of constraints is defined on activation values of the incoming and outgoing edges of a node depending on its types.\\
{\bf Kidney nodes (K):} This type of node can either have only an outgoing edge activated, or all the edges inactive, because each kidney can be given only to one patient or to no one.\\
{\bf Pair nodes (P):}. This type of node can either have only an incoming edge activated, or one incoming and one outgoing edge activated or all the edges inactive. This because a pair cannot donate a kidney without receiving another kidney, but can receive and not donate because in this case the kidney of the donor will be donated to the patients in the deceased-donor waiting list.\\
{\bf Immunized patient nodes (UT):} This type of node can have either only an incoming edge active, or all the edges inactive. This because each patient can either receive only one kidney or no kidney.

We can translate the constraints described above more formally as:
$$
\begin{cases}
v_e \in \{0, 1 \} ~~~ \forall e \in E &\text{activation}
\\
\sum_{e \in E_X} v_e \in \{0, 1 \} ~~~ \forall X \in K & \text{K nodes}
\\
\sum_{e \in E_X^-} v_e - \sum_{e \in E_X^+} v_e \in \{0, 1 \} ~~~ \forall X \in P & \text{P nodes}
\\
\sum_{e \in E_X} v_e  \in \{0, 1,2 \} ~~~ \forall X \in P & \text{P nodes}
\\
\sum_{e \in E_X} v_e \in \{0, 1 \} ~~~ \forall X \in UT & \text{UT nodes}
\\
\end{cases}
$$

In our experiments we considered also an additional constraint over the compatibility graph: if we include in our pool also kidneys that we would allocate to an immunized patient (UT) in the deceased donor waiting list, then we want to preserve the number of $UT$ patients that receive an organ. This because we want to prioritize this category and not create a disadvantage.
The formulation of this constraint is:
$$
\sum_{X \in UT} (\sum_{e \in E_X^-} v_e) - \sum_{X \in K} (\sum_{e \in E_X^-} v_e) \geq 0
$$




\subsection{Objective function}

At each iteration of the algorithm (either cycle or chain detection), we maximize the following objective function: $\sum_{e \in E} v_e $, that maximize the number of active edges in the graph. This corresponds to maximizing the number of transplants. It is important to notice that a priority over the UT patients respect to normal patients in the waiting list is implicitly encoded in the graph construction.
Once obtained a solution (a set of active edges), the patients/pairs that will be transplanted correspond to the subset of nodes that are touched by active edges. We then remove this nodes/patients for the next iteration of the procedure.

\subsection{Main procedures}

We describe now, formally, the two different procedures:
\\
{\bf Cycle detection:} This happen when at time $t$ a new pair joins the pool. We build a compatibility graph using only the pair nodes currently part of the pool $P(t)$. On the compatibility graph, we then maximize the length of a cycle involving the new pair, if exist. To this purpose we create another graph, namely the {\it cycle graph}, in which each node corresponds to a cycle of length 2 or 3 (this is a common assumption in this scenarios, since all the transplants in the cycle have to be performed simultaneously). The edges of the cycle graph connect two nodes representing cycles with at least one node in common in the compatibility graph. We solve this optimization problem maximizing the cardinality of a coloring problem over the cycles graph.
\\
{\bf Chain detection:} This happen when at time $t$ a new deceased-donor kidney arrives to the waiting list.  
We build a compatibility graph using the pair nodes, the UT nodes currently part of the pool $P(t)$ and and the kidney node. Here we look for the maximum chain of nodes starting from the kidney. We solve this problem as a single source longest path problem\footnote{Note that we assume the pool $P(t)$ to be acyclic, since all the transplant in the cycles are performed in previous steps}.

\subsection{Assumptions}

The implementation of the algorithm that we used for our experiments is based on two assumptions/restrictions.
The first assumption is that, during the chain detection phase, the compatibility graph is assumed to be acyclic since all the transplant in the cycles are performed in previous steps. This is a reasonable assumption for our study, but in reality many cycles are not performed since patients involved prefer to wait to obtain a better graft. Thus the compatibility graph could contain some cycles.
The second restriction is that during the cycle detection our algorithm focus only on cycles that involve the new pair. This is a reasonable assumption for our study since we are operating in a small pool (described in the Experiments Section). In practice if we consider a bigger pool (for example at national level), the introduction of new pairs is done monthly. This implies that at that time $t$ usually more than one pair is introduced to the pool. In this case we are not interested on maximizing the length of a single cycle but on maximizing the total number of transplants. That means we are looking for the set of cycles that maximally cover the graph without intersections.

We also implemented a different version of the algorithm for a general use, that supports also the removal of the two restriction described above. The new version of the algorithm is a reformulation of the maximization problem in integer linear programming (ILP). The objective function to maximize is the sum of active edges and the constrains correspond to a ILP version of the constraints on the compatibility graph described at the beginning of the section.
We implemented the algorithm in Python using the open-source library {\em lp\_solve}. 


\begin{table*}[!ht]
\begin{center}
\begin{tabular}{|l|l|l|l|l|l|l|} \hline
\rowcolor{gray!50} Characteristics & Kidneys & UT patients & NT donors & NT patients & DS donors & DS patients \\ 
SEX F/M &  $30$/$39$ & $14$/$21$ &  $11$/$7$ & $8$/$8$ &  $22$/$8$ & $13$/$17$\\
\hline
AGE (mean) & $48$ & $47$ & $54.34$ & $47.12$ & $48.63$ & $42.07$\\
\hline
BLOOD type A/B/AB/0 & $30$/$11$/$4$/$24$ & $13$/$2$/$1$/$19$ & $7$/$4$/$0$/$7$ & $4$/$3$/$0$/$9$ & $15$/$2$/$6$/$7$ & $8$/$5$/$1$/$16$\\
\hline
PRA max (mean) & - & $58.43$ & - & $56.56$ & - & - \\
\hline
\end{tabular}
\caption{Pool characteristics}
\label{table:data}
\end{center}
\end{table*}

\subsection{Examples}

We now show two examples, one for the {\em cycle detection} step of the procedure in Example \ref{ex:reni_cycles} and one for the {\em kidneys processing} step in Figure \ref{ex:reni_chain}.

\begin{example}[Cycle detection]\label{ex:reni_cycles}
In the  {\em cycle detection} example (Figure \ref{fig:reni_cycles}) we have four pairs $P.0$, $P.1$, $P.2$ and $P.3$. 
The patient in $P.0$ is compatible with the donor in $P.1$;
the patient in is $P.1$ compatible with the donors in $P.0$ and $P.2$;
the patient in is $P.2$ compatible with the donors in $P.0$ and $P.1$; and the patient in is $P.3$ compatible with the donor in $P.2$.
We can see that the algorithm detects the cycle with the maximal number of transplants (the edges in red): between nodes $P.0$, $P.1$ and $P.2$.

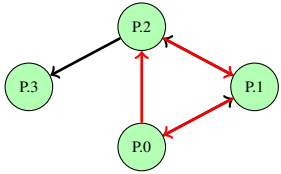
\begin{figure}[h!]
\tiny
\begin{center}
\begin{tikzpicture}[scale=1]
\tikzstyle{every node}=[draw,shape=circle,fill=green!30,minimum size=3em];
\node (0D/R) at (1.5,0) {P.0};
\node (1D/R) at (3,0.8) {P.1};
\node (3D/R) at (0,0.8) {P.3};
\node (2D/R) at (1.5,1.6) {P.2};
\draw [->,line width=1pt]  (0D/R) -- (1D/R);
\draw [->,red,line width=1pt]  (1D/R) -- (0D/R);
\draw [->,line width=1pt]  (1D/R) -- (2D/R);
\draw [->,red,line width=1pt]  (2D/R) -- (1D/R);
\draw [->,line width=1pt]  (2D/R) -- (3D/R);
\draw [->,red,line width=1pt]  (0D/R) -- (2D/R);
\end{tikzpicture}
\caption{Example \ref{ex:reni_cycles}: cycle detection.}
\label{fig:reni_cycles}
\end{center}
\end{figure}

\end{example}

\begin{example}[Chain detection]\label{ex:reni_chain}
In the  {\em kidneys processing} example in Figure \ref{fig:reni_chain}, we have four nodes: tree pairs $P.0$, $P.1$ and $P.2$; one kidney from a deceased donor $K$; and one UT patients $UT$. 
The patient in $P.0$ is compatible only kidney from a deceased donor $K$;
the patient in is $P.1$ is compatible with the donor in $P.0$ and kidney from a deceased donor $K$;
the patient in is $P.2$ is compatible with the donor in $P.1$ ; and the patient in $UT$ is compatible with the donors in $P.0$  and $P.1$.
We can see that the algorithm detects the longest chain that maximize the number of transplants that starts from $K$ then reach in order $P.0$, $P.1$ and ends in $UT$.


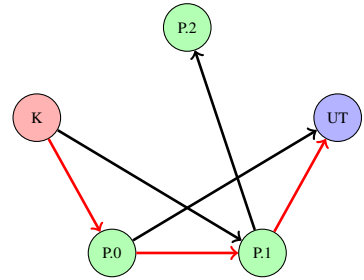
\begin{figure}[h!]
\tiny
\begin{center}
\begin{tikzpicture}[scale=1]
\tikzstyle{every node}=[draw,shape=circle,fill=blue!30,minimum size=3em];
\node (2R) at (4,1.8) {UT};
\tikzstyle{every node}=[draw,shape=circle,fill=green!30,minimum size=3em];
\node (0D/R) at (1,0) {P.0};
\node (1D/R) at (3,0) {P.1};
\node (2D/R) at (2, 3) {P.2};
\tikzstyle{every node}=[draw,shape=circle,fill=red!30,minimum size=3em];
\node (4D) at (0,1.8) {K};
\draw [->,red,line width=1pt]  (4D) -- (0D/R);
\draw [->,line width=1pt]  (4D) -- (1D/R);
\draw [->,line width=1pt]  (0D/R) -- (2R);
\draw [->,red,line width=1pt]  (0D/R) -- (1D/R);
\draw [->,line width=1pt]  (1D/R) -- (2D/R);
\draw [->,red,line width=1pt]  (1D/R) -- (2R);
\end{tikzpicture}
\caption{Example \ref{ex:reni_chain}: chain detection.}
\label{fig:reni_chain}
\end{center}
\end{figure}

\end{example}
\section{Experiments}

While the cycles detection procedure is well studied in the literature from an ethical point of view, the chain detection starting from a deceased donor is a novel method. Our sequential algorithm for chain detection works as follows on our pool: when an organ from a deceased donor becomes available, we compute all the chains that could be generated by assigning this organ to some recipient with an incompatible living donor. Among all chains, we perform the longest one and we remove all recipients and corresponding donors from the pool. When the following organ becomes available we iterate the procedure with the remaining pool.
The recipient of the deceased donor organ could be any of the 
patients with an incompatible donor who were not transplanted during the period. Receiving a transplant, even from a deceased donor, is a sure gain for those patients who were not transplanted, but one may argue that this policy would subtract organs from the pool available for waitlisted candidates. The organ harvested from a living donor of the last component of the chain will terminate to a deceased donor waitlisted patient; therefore waitlisted patients without a potential living donor do not suffer due to the introduction of this allocation procedure - because the expected graft survival of a living donor kidney is on average higher than a deceased donor one. Nevertheless, there has to be some equity warranted for the waitlisted candidates with a lower probability of finding a compatible organ. To this end, we add two constraints to our algorithm. First, deceased donor organs that were directly allocated to recipients of unlikely transplantability  recipients were excluded from the algorithm, and we only consider organs allocated to patients in the standard waiting list. Second, anytime there exists more than one living donor chain with the maximal length, we select the one, if any, that ends with an UT recipient.

All the experiments were conducted using a custom Python code on a machine with an Eight-Core Intel(R) Xeon(R) 2.40GHz with 256GB of RAM.

\subsection{Data}

From January 2012 to December 2014 at the Kidney and Pancreas Transplantation Unit of the Padua University-Hospital 358 single kidney transplants (KT) were performed. Among them, 251 KT were done with grafts from deceased donors and 107 from living donors. The living donors transplantation included: 77 AB0-compatible pairs and 30 desensitized recipients (22 AB0 incompatible pairs  and 8 recipients with donor-specific anti-HLA antibodies). During the same span, 16 incompatible pairs were evaluated in the centre and were enrolled in the KPE program and/or listed for deceased donor kidney transplantation and could not be transplanted within the three-year period (in 2 cases a recipient had 2 willing donors). 
Among the DD kidneys that were allocated to this centre in the relevant span, we only consider organs of from deceased donors of "high quality", that is with an expected graft survival comparable to that of a living donor, according to the medical literature (age $<$ 60 years, absence of comorbidities such as hypertension and diabetes, calculated creatinine clearance $>$ 60 ml/min, absence of proteinuria). We also excluded organs that were assigned to UT patients .
UT recipients were defined based on the classification of The Nord Italia Transplant Program (NITp), which considers UT recipients those patients on the waiting-list for more than 5 years or on dialysis for more than 7 years. During the period of study, (2012-2014), 35 UT recipients of unlikely transplantability were listed at our Centre and did not receive any organ.
Summing up, data used to run our retrospective studies are the following: 16 NT pairs, 30 DS pairs, 35 UT recipients, 69 DD "high-quality" kidneys.

The population of DS pairs benefit from the participation to the program because recipients of these pairs could avoid debilitating and costly desensitization procedures. However, these patients should be considered differently than patients who were not transplanted to ensure that they are willing to enter into the program. To this end, we impose two constraints to the algorithm. First, for desensitised patients (DS patients), the advantage of receiving a kidney from a compatible donor in a KPE fashion should not prolong their waiting time on dialysis: therefore, we grant the availability of these pairs in the pool within 6 months from the date of their actual transplant; after this date recipients are desensitized and get the organ of their willing donor. Second, DS patients can only receive an organ from a living donor and not from a deceased donor, and therefore they cannot be the first recipient of a living donor chain initiated by a deceased donor organ. 

For a proper allocation of the organs, immunological variables of donors and recipients were taken into account (e.g. HLA typing, blood type, etc.).
The statistics on the distribution
\footnote{The number of donors of the NT pairs could be greater than the number  of patients because some patients have multiple donors. This does not happen for the DS pairs since these pairs have already been transplanted with their donor that is therefore unique.} 
of sex, age, blood type and PRA max of the above described pool are summarized in Table \ref{table:data}.


\subsection{Experiments' Design}


\begin{table*}[!h]
\begin{center}
\begin{tabular}{|l|l|l|l|l|l|l|l|l|} \hline
\rowcolor{gray!60} Number of: & 
-C- - & -C-$0$ & -CD- &   -CD$0$  &  C-D- & CCD- & CCD$0$ & CCD$0^+$  \\ 
\hline
\hline
NT patients who received an organ &
$8$/16 & $8$/16  & $12/16$ & $9/16$ & $4/16$ & $12$/16 & $10$ /16& $7/16$ \\ \hline
DS patients who avoided desensitization &
    0/16   & 0/16      & $10/16$ & $4$/16  & $6$/16& $10$/16& $10$/16 & $6/16$ \\ \hline
UT patients who received an organ &
$11/35$& $6/35$  & $9/35$   & $7/35$  & 0/35     & $9/35$    & $5/35$  & $3/35$\\ \hline
Deceased donor kidneys that are used &
$15$/69& $7/69$  & $14$/69 & $8$/69 &0/69     & $14/69$  & $5/69$  & $3/69$ \\ \hline
Living donor kidneys returned to the waiting list &
$4$  & $1$  & $5 $  & $1$  & 0     & $5$    & $0$  & $0$\\ 
\hline
\end{tabular}
\caption{Results of the performed experiments.}
\label{table:experiments_results}
\end{center}
\end{table*}


We performed a set of experiment varying a set of 4 parameters. Each experiment is labeled with four letters filling a slot of four fields, each one corresponding to the presence of the corresponding parameter. If a parameter is missing we denote it with the hyphen symbol: $-$.
We describe now the four ordered fields:\\
{\bf First field (C):} In these experiments we perform cycles detection when a new pair enter the pool. We consider the sequence of entering of pairs to the waiting list. Any time a new pair join the program we ran the algorithm searching the longest cycle involving the new pair. In the experiments without the 'C' as first field we do not look for cycles but only for chains of donations.\\
{\bf Second field (C):} In these experiments we perform chain detection when a new deceased-donor kidney arrive to the waiting list. We consider the sequence of deceased donor organs that were available in the relevant time span to initiate living donor chains and any time a new organ was available we ran the algorithm searching the longest living donor chain initiated by this organ. In the experiments without the 'C' as second field we do not search for chains but only cycles of donations.\\
{\bf Third field (D):} We add to the pool the pairs of DS patients (NT pairs are always in the pool). The algorithm works in a similar fashion to the previous one, but in addition, when an incompatible pair requiring desensitization joins the pool, we immediately check if it is possible to create a two or three-way exchange among the pairs who are in our pool at that moment. The population of DS pairs benefit from the participation to the program because recipients of these pairs could avoid debilitating and costly desensitization procedures. However, these patients should be considered differently than patients who were not transplanted to ensure that they are willing to enter into the program. To this end, we impose two constraints to the algorithm. First, a DS patient and her donor remain in the pool for at most six months. During this span, they are scrutinized in each algorithm iteration: that is when either a new deceased donor organ becomes available, or a new incompatible pair requiring desensitization joins the pool. After six months, if no exchange was possible, they will leave the pool, and the transplant is performed using desensitization between the pair. Second,  a DS patient can only receive an organ from a living donor and not from a deceased donor, and therefore cannot be the first recipient of a living donor chain initiated by a deceased donor organ. In the experiments without the 'D' as third field we do not introduce in the pool of desensitized pairs, but only NT pairs.\\
{\bf Fourth field (0):} Since the kidneys with $0$ blood type are more favorites (since compatible with all the other blood types) we want to preserve the number of  $0$-type organs given to the deceased-donor waiting list. To do this, when we receive a $0$-type kidney we only allow the chain either to finish with a UT-patient or returning a $0$-type kidney to the deceased-donor waiting list. In the experiments without the `$0$' as last field the chains can end with any patient or pair, without any restriction on their blood type.

We provide now some example of experiments labels associated with their meaning.
\begin{example}
For example the experiment -C- - represents the dynamic computation of only chains on a pool composed by: NT pairs, a deceased-donor kidney and UT patients.
The experiment CCD$0$ instead represent the dynamic computation of chains and cycles on a pool composed by: NT pairs, DS pairs, a deceased-donor kidney and UT patients with the $0$-type preservation constraint.
\end{example}
It is important to notice that not all the combinations of these four parameters is feasible. For example all the combinations with the first two fields both empty it is not possible, since we want to perform at least one procedure between chain detection or cycle detection.

Moreover, in our experiments we considered also a stronger version of the $0$-type preservation constraint, that will be denoted as {\bf $0^+$}: as in the $0$-type preservation constraint with the addition that if the deceased-donor kidney has $0$ blood type then we allow the chain either to finish with a UT-patient with $0$ blood type or returning a $0$ blood type kidney to the deceased-donor waiting list. 


\subsection{Results and Discussion}
The results of the retrospective simulations are summarized in Table \ref{table:experiments_results}.

Analyzing for example the experiment $-C-0$, we notice that given a cohort of 16 incompatible pairs, and a pool of 69 standard deceased-donors allocated to the Padua Transplant Centre, it turns out that by using 7 grafts from DDs to start a chain, it was theoretically possible to transplant 50\% of the patients who would have not been otherwise able to receive a transplant in a time span of three years. This means that only 10\% of the entire pool of standard grafts available was utilized to enter the program.
 Moreover, in most of the cases (6 out of 7) the chains ended to a UT recipient who therefore received a living donor kidney instead of a deceased-donor organ.

However, it is noteworthy that, differently from in list exchanges, patients participating in a chain of donations receive a kidney before her/his donor donates an organ to another patient. Moreover, for each incompatible pair, who can start a chain of donations, it is possible to compute in advance (in offline fashion) which is the optimal chain to perform, making the logistics easier.

 The usefulness of the proposed program results undeniable, since it increases the overall number of kidneys available for transplantation and, consequently, the aggregate quality and quantity of life of end-stage-renal-disease patients.
Our procedure has been approved  and a first trial at the local level of a KP chain initiated by a DD has been performed on March 14, 2018 and two more chains respectively of three and four length are currently  performed at national level showing the  benefits of the program also in practice.

\section{Conclusion and Future work}

In this paper we have analyzed the efficiency of a novel allocation procedure for kidney transplant: we have proposed a first method based on the idea of considering both cycles of exchanges and chains of exchanges which start from a deceased donor kidney. 
We have implemented the procedure and have tested it in a retrospective fashion, using historical data from Padua's transplantation center. The results show that we increase both the number of incompatible pairs that are transplanted and the number of highly-sensitized (UT) patients that received a donation (that normally have a very low probability to receive a compatible kidney). Moreover we have avoided desensitization by the majority of the patient/donor pairs that were transplanted with desensitization. 

An immediate next step for our work would be the extension the program to more than one single centre. This would likely lead to better results, since the pool would be bigger increasing the probability of compatibility. 

A possible further expansion of this program that combines living and deceased donors  is represented by the inclusion of compatible pairs to the pool of participants, encouraging them to enroll in the program with the promise of a gain in terms of expected quality of the organ with respect to the intended donor.

We plan also to consider the patients' preferences in order to minimize the number of matching rejections. For instance we could include the patients' preferences about the location of the hospital (to avoid rejections for logistical problems) or estimate the risk aversion of each patient and the corresponding donor (to minimize the possibility of withdraws due to the non-simultaneously of the transplants).

\newpage
\bibliographystyle{aaai}

\end{document}